%
% 2dF Beta Nature Article
%
% a plain TeX document - God rot LaTeX
%
\mag 1200
%
% local macros
%
\vsize=9.6truein \hsize=6.3truein
\parskip =2ex plus .5ex minus .1ex
\parindent 3em
\font\sc = cmcsc10
\font\bfnew= cmbx7 at 10pt
\font\smallit= cmti10 at 7pt
\font\big = cmbx10
\font\bbig = cmbx10 scaled \magstep2
\textfont\bffam=\bfnew
\def\bf{\fam\bffam\bfnew}
\def\nref#1{$^{\bf #1}$}
\def\ref#1{\parskip=0pt\par\noindent$^{\bf #1}$\hangindent\parindent
    \parskip =2ex plus .5ex minus .1ex}
\def\eol{\hfill\break}
\def\gs{\mathrel{\lower0.6ex\hbox{$\buildrel {\textstyle >}
 \over {\scriptstyle \sim}$}}}
\def\ls{\mathrel{\lower0.6ex\hbox{$\buildrel {\textstyle <}
 \over {\scriptstyle \sim}$}}}

\def\kms{{\,\rm km\,s^{-1}}}
\def\kmsmpc{{\,\rm km\,s^{-1}Mpc^{-1}}}
\def\hompc{{\,h\,\rm Mpc^{-1}}}
\def\mpcoh{{\,h^{-1}\,\rm Mpc}}
\newfam\sodfam
\def\sodyou{\fam\sodfam\sodfont}
\font\sodfont = cmmib10
\textfont\sodfam=\sodfont
\mathchardef\chbbeta="710C
\def\bbeta{{\sodyou\chbbeta}}
\def\fig#1#2#3#4{\midinsert\vbox{
 \epsfxsize=0.01\hsize
 \multiply\epsfxsize by #3
 \medskip
 \centerline{#4}
 \smallskip
% \narrower
 \advance\leftskip by 1.5em
 \advance\rightskip by 1.5em
 \lineskiplimit=-10pt
 \baselineskip = 0.45truecm\noindent{\bf Figure #1}
 \quad {\rm#2}\par}\endinsert
}
\overfullrule 0pt

\input epsf
\strut
\vglue -1cm
\noindent
{{\it Nature\/} {\bf 410}, 169--173 (2001)}
\bigskip
{\bbig
\noindent
A measurement of the cosmological mass density 
from \hfill\break
clustering in the 2dF Galaxy Redshift Survey}

\bigskip

\noindent
{\big
John A. Peacock$^{1}$,
Shaun Cole$^2$, 
Peder Norberg$^2$,
Carlton M. Baugh$^2$,
Joss Bland-Hawthorn$^3$,
Terry Bridges$^3$, 
Russell D. Cannon$^3$, 
Matthew Colless$^4$, 
Chris Collins$^5$, 
Warrick  Couch$^6$, 
Gavin Dalton$^7$,
Kathryn Deeley$^6$, 
Roberto De Propris$^6$,
Simon P. Driver$^8$, 
George Efstathiou$^9$, 
Richard S. Ellis$^{9,10}$, 
Carlos S. Frenk$^2$, 
Karl Glazebrook$^{11}$, 
Carole Jackson$^4$,
Ofer Lahav$^9$, 
Ian Lewis$^3$, 
Stuart Lumsden$^{12}$, 
Steve Maddox$^{13}$,
Will J. Percival$^1$,
Bruce A. Peterson$^4$, 
Ian Price$^4$,
Will Sutherland$^{7,1}$,
Keith Taylor$^{3,10}$}

%J.A. Peacock$^{1}$,
%S. Cole$^2$, 
%P. Norberg$^2$,
%C.M. Baugh$^2$,
%J. Bland-Hawthorn$^3$,
%T. Bridges$^3$, 
%R.D. Cannon$^3$, 
%M. Colless$^4$, 
%C.A. Collins$^5$, 
%W.J. Couch$^6$, 
%G. Dalton$^7$,
%K.E. Deeley$^6$, 
%R. De Propris$^6$,
%S.P. Driver$^8$, 
%G. Efstathiou$^9$, 
%R.S. Ellis$^{9,10}$, 
%C.S. Frenk$^2$, 
%K. Glazebrook$^{11}$, 
%C. Jackson$^4$,
%O. Lahav$^9$, 
%I.J. Lewis$^3$, 
%S.L. Lumsden$^{12}$, 
%S.J. Maddox$^{13}$,
%W.J. Percival$^1$,
%B.A. Peterson$^4$, 
%I. Price$^4$,
%W. Sutherland$^{7,1}$,
%K. Taylor$^{3,10}$}

\eol
{\smallit
$^{1}$Institute for Astronomy, University of Edinburgh, Royal Observatory, Blackford Hill, Edinburgh EH9 3HJ, UK \eol
$^2$Department of Physics, University of Durham, South Road, Durham DH1 3LE, UK \eol
$^3$Anglo-Australian Observatory, P.O.Box 296, Epping, NSW 2121,  Australia\eol  
$^4$Research School of Astronomy \& Astrophysics, The Australian  National University, Weston Creek, ACT 2611, Australia \eol
$^5$Astrophysics Research Institute, Liverpool John Moores University, Twelve Quays House, Birkenhead, L14 1LD, UK \eol
$^6$Department of Astrophysics, University of New South Wales, Sydney, NSW 2052, Australia \eol
$^7$Department of Physics, University of Oxford, Keble Road, Oxford OX3RH, UK \eol
$^8$School of Physics and Astronomy, University of St Andrews, North Haugh, St Andrews, Fife, KY6 9SS, UK \eol
$^9$Institute of Astronomy, University of Cambridge, Madingley Road, Cambridge CB3 0HA, UK \eol
$^{10}$Department of Astronomy, Caltech, Pasadena, CA 91125, USA \eol
$^{11}$Department of Physics \& Astronomy, Johns Hopkins University, Baltimore, MD 21218-2686, USA \eol
$^{12}$Department of Physics, University of Leeds, Woodhouse Lane, Leeds, LS2 9JT, UK \eol
$^{13}$School of Physics \& Astronomy, University of Nottingham, Nottingham NG7 2RD, UK \eol
}

\medskip

\hrule height0.05cm

\bigskip

\noindent
{\bf
The large-scale structure in the distribution of galaxies
is thought to arise from the gravitational
instability of small fluctuations in the
initial density field of the universe. A
key test of this hypothesis is that superclusters of galaxies in the
process of formation should generate systematic
infall of other galaxies. This would be evident in the pattern 
of recessional velocities, causing an anisotropy in
the inferred spatial clustering of galaxies.
Here we report a precise measurement of this clustering,
using the redshifts of more than 141,000 galaxies from
the two-degree-field galaxy redshift survey.
We determine the parameter
$\bf\bbeta\equiv \Omega^{0.6}/b = 0.43 \pm 0.07$,
where $\bf\Omega$ is the total mass-density parameter and $\bf b$ is a
measure of the `bias' of the
luminous galaxies in the survey. Combined with the
anisotropy of the cosmic microwave background, our results 
favour a low-density universe with $\bf\Omega\simeq 0.3$.
}

\vfill\eject

\noindent
Hubble showed in 1934 that the pattern of galaxies
on the sky is non-random\nref{1}, 
and successive years have seen ever more ambitious
attempts to map the distribution of
visible matter on cosmological scales. 
In order to obtain a three-dimensional picture, redshift
surveys use Hubble's law, $v=H_0 r$, to infer approximate
radial distances to a set of galaxies.
The first major surveys of this sort took place in the early
1980s\nref{2,3,4,5}, and were limited to a few thousand
redshifts, owing to the limited speed of single-object
spectroscopy. In the 1990s, redshift surveys were extended to much
larger volumes by a `sparse sampling' strategy\nref{6}.
These studies\nref{7,8} established that
the universe was close to uniform on scales above
about $100\mpcoh$ ($h\equiv H_0/100\kmsmpc$), but with a complex nonlinear
supercluster network of walls, filaments and voids on smaller scales.

The origin of this large-scale structure is one of the key
issues in cosmology. A plausible assumption is that structure
grows via gravitational collapse of density fluctuations that
are small at early times -- but it is vital to test this idea.
One important signature of gravitational instability is that 
collapsing structures should generate `peculiar'
velocities, ${\bf \delta v}$, which distort the uniform Hubble expansion.
We measure a redshift, $z$, which combines
Hubble's law with the radial component of these peculiar
velocities:
$cz \simeq H_0 r + {\bf \delta v\cdot \hat r}$. 
The apparent density field seen in a
redshift survey is thus not a true three-dimensional picture, 
but this can be turned to our advantage. The
redshift-space  distortions have a characteristic form,
whose detection can both verify the general idea that
structure forms by gravitational instability,
and also measure the density of the universe.
The present paper presents measurements of this effect,
based on a new large redshift survey.

\noindent
{\big The 2dF Galaxy Redshift Survey}\eol
New-generation large redshift surveys are
made feasible by multiplexed fibre-optic spectroscopy, and
the most advanced facility of this sort is the
2-degree Field, mounted at the prime focus of the
Anglo-Australian Telescope\nref{9,10}, which allows 400
spectra to be measured simultaneously. 
The 2dFGRS\nref{11,12} was designed to use
this instrument to measure the redshifts of 250,000
galaxies, to a blue magnitude limit of 
$b_{\scriptscriptstyle\rm J}=19.45$ (corrected for extinction
by dust in the Milky Way).
The galaxies were selected from an updated version of the
APM catalogue\nref{13}, which is based on scans of
photographic plates taken with the UK Schmidt Telescope.
Survey observations began in 1998, and should finish at
the end of 2001. At the time of writing, redshifts have
been obtained for 141,402 galaxies.
The sky coverage of the 2dFGRS consists of strips in
the Northern and Southern Galactic Poles ($75^\circ \times 7.5^\circ$ in
the NGP; $75^\circ \times 15^\circ$ in the SGP), plus a number
of outlying random fields. Coverage is now sufficiently
extensive that near-complete thin slices through the galaxy
distribution may be constructed, as shown in Fig. 1.
This image illustrates well the median depth of the
survey: approximately $z=0.11$. Beyond this point, the survey is
sensitive only to the more luminous galaxies, and
the comoving density falls rapidly. Nevertheless, the
survey volume (including other regions not shown) is
more than adequate for an accurate determination of the
statistical properties of galaxy clustering.

\fig{1}
{The distribution of galaxies in part of the 2dFGRS, drawn from 
a total of 141,402 galaxies:
the slices are $4^\circ$ thick, centred at declination
$-2.5^\circ$ in the NGP (left) and $-27.5^\circ$ in the SGP (right).
Not all 2dF fields within the slice have been observed
at this stage, hence there are weak variations of the
density of sampling as a function of right ascension.
To minimise such features, the slice thickness increases
to $7.5^\circ$ between right ascension $13.1^h$ and $13.4^h$.
This image reveals a wealth of detail, including
linear supercluster features, often nearly perpendicular
to the line of sight. The interesting question
to settle statistically is whether such transverse features
have been enhanced by infall velocities.}
{100} 
{\epsfbox[57 232 552 559]{slices.eps}}

\fig{2}
{The redshift-space correlation function for the 2dFGRS,
$\xi(\sigma, \pi)$, plotted as a function of transverse 
($\sigma$) and radial ($\pi$)  pair
separation. The function was estimated by counting pairs
in boxes of side $0.2\mpcoh$
(assuming an $\Omega=1$ geometry), and then smoothing with a
Gaussian of rms width $0.5 \mpcoh$. To illustrate deviations
from circular symmetry, the data from the first quadrant are
repeated with reflection in both axes.
This plot clearly displays redshift distortions, with
`fingers of God' elongations at small scales and the coherent
Kaiser flattening at large radii.
The overplotted contours show model predictions
with flattening parameter $\beta\equiv \Omega^{0.6}/b = 0.4$
and a pairwise dispersion of $\sigma_p = 400 \kms$.
Contours are plotted at $\xi = 10,5,2,1,0.5,0.2,0.1$.\eol
\strut\hglue\parindent
The model predictions assume that the redshift-space
power spectrum ($P_s$) may be expressed as a product of
the linear Kaiser distortion and a radial convolution\nref{14}:
$P_s({\bf k}) = P_r(k) \, (1+\beta \mu^2)^2 \, (1+ k^2 \sigma_p^2 \mu^2 /2 H_0^2)^{-1}$,
where $\mu={\bf \hat k\cdot \hat r}$, and $\sigma_p$ is 
the rms pairwise dispersion of the random component of the 
galaxy velocity field.
This model gives
a very accurate fit to exact nonlinear simulations\nref{15}.
For the real-space power spectrum, $P_r(k)$, we take the
estimate obtained by deprojecting the angular clustering in the APM
survey\nref{13, 16}. This agrees very well with estimates that can
be made directly from the 2dFGRS, as will be discussed elsewhere.
We use this model only to estimate the scale dependence
of the quadrupole-to-monopole ratio
(although Fig. 2 shows that it does 
match the full $\xi(\sigma,\pi)$ data very well).
}
{70}
{\epsfbox[0 0 390 394]{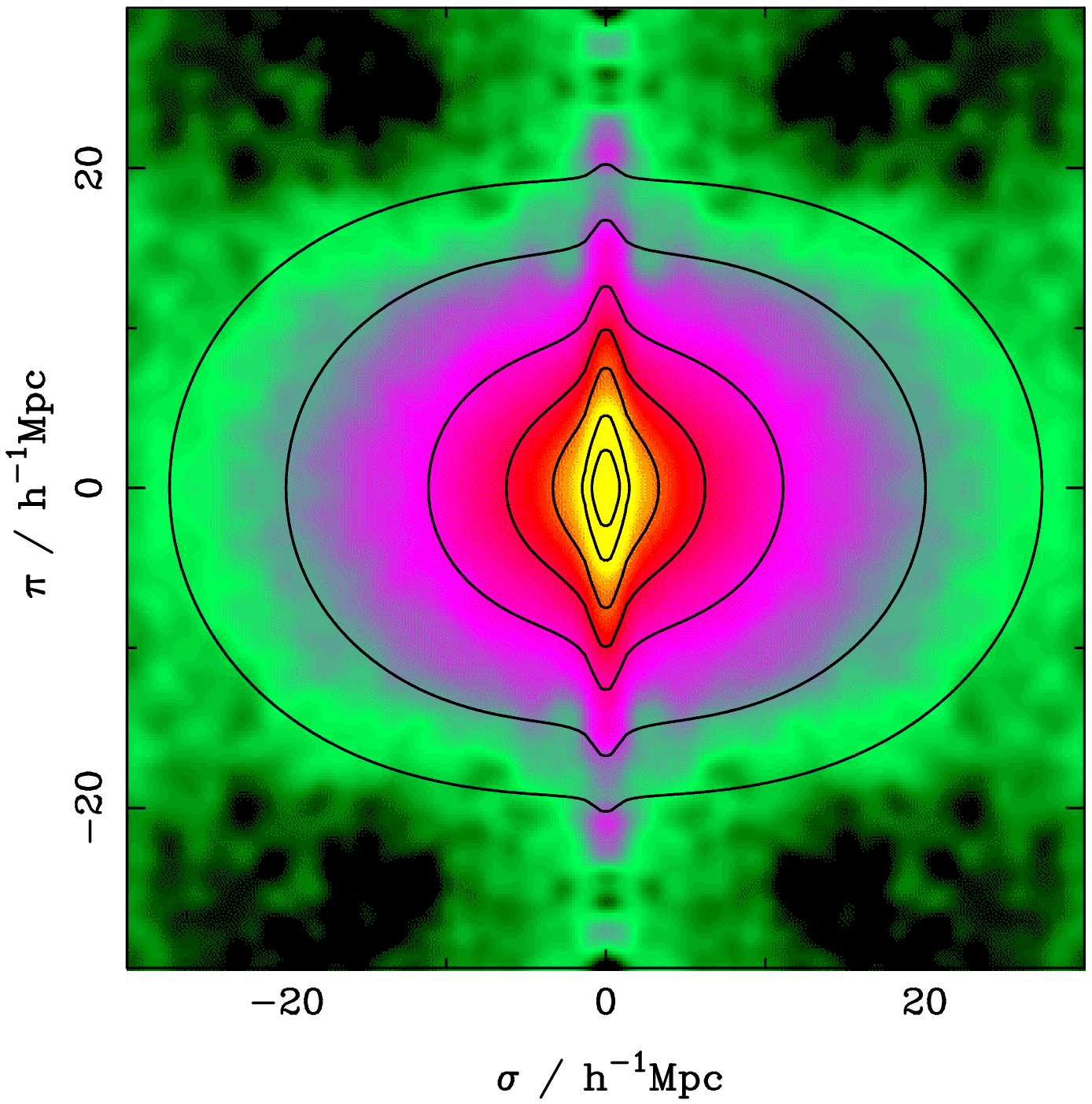}}

\noindent
{\big Galaxy correlations in redshift space}\eol
The simplest statistical indicator of 
peculiar velocities in cosmological structure is the two-point correlation
function, $\xi(\sigma,\pi)$. This measures the
excess probability over random of finding a pair of galaxies with
a transverse separation $\sigma$ and a line-of-sight
separation $\pi$. In an isotropic universe, this
function should be independent of direction, but this is
not true in redshift space. Transverse
separations are true measures of distance, but apparent radial
separations are distorted by peculiar velocities. 
This redshift-space anisotropy should
cause two characteristic effects, operating respectively
on small and large scales. On small scales, random orbital
velocities within galaxy groups cause an apparent radial
smearing, known as `fingers of God'. Of greater interest is
the large-scale effect; if cosmological structure forms
via gravitational collapse, there should exist coherent
infall velocities, and the effect of these is to cause an
apparent {\it flattening\/} of structures along the 
line of sight. The general existence of redshift-space distortions 
was recognized in the first redshift surveys\nref{2,3,4}, but the
first comprehensive analysis of the phenomenon
was performed by Kaiser\nref{17},
who showed that they could be used to measure the quantity
$$
\beta \equiv \Omega^{0.6} / b,
$$
where $\Omega$ is the cosmological mass density parameter and $b$
is the bias parameter that relates the relative density
fluctuations of the galaxies and of the total mass:
$$
\left.{\delta \rho \over \rho}\right|_{\rm galaxies} = 
b \left.{\delta \rho \over \rho}\right|_{\rm mass}.
$$
The presence of bias is an 
inevitable consequence of the nonlinear nature
of galaxy formation, and the relation between mass and galaxy
tracers is complex\nref{18,19,20}. However, there are good theoretical reasons to
expect that $b$ can indeed be treated as a constant on
large scales, where the density fluctuations are linear\nref{21, 22}.
Redshift-space distortions have thus been seen as an important
method for weighing the universe\nref{23, 24}. 
To date, a number of papers have made significant detections of the
Kaiser effect\nref{18,25,26}, but the 2dFGRS is the first survey that is large enough
for the effect to be studied in detail.

In order to estimate $\xi(\sigma,\pi)$,
we follow standard methods\nref{27,28} that compare
the observed count of galaxy pairs with the count estimated
using a random distribution that obeys the same selection
effects in redshift and sky position. These selection effects
are well defined, but complex: the survey is tessellated into a
pattern of `sectors' defined by the overlap of 
the $2^\circ$ diameter survey tiles, whose positions are
chosen adaptively with the aim of being able to place a fibre
on $>95$\% of the galaxies in the input catalogue. At the present
intermediate stage of the survey, many tiles remain
to be observed, and some regions of the survey presently contain redshifts
for $<50$\% of the galaxies. Furthermore, the
spectroscopic success rate (redshifts per allocated fibre)
is $>95$\% in good conditions, but can fall to $\simeq 80$\%
in marginal weather. We have implemented a number of
independent algorithms for
estimating the resulting survey selection effects, and are confident that
we can measure the galaxy correlations robustly out to
a separation of $25\mpcoh$.
For example, the redshift distribution in sectors with low
spectroscopic completeness is biased to low redshifts, but it
makes no significant difference whether or not we correct for this, or indeed
whether the low-completeness regions are simply excised.
In addition to allowing for survey completeness, it is necessary
to give higher weight to regions with a low sampling density, to achieve
the optimum balance between cosmic variance and shot noise\nref{6}.
In practice, we have chosen to truncate the analysis at
a maximum redshift of $z=0.25$. Within this volume, the exact
optimum weight per galaxy varies very nearly as the reciprocal of the
number density, so that all volume elements receive approximately
equal weight.
The redshift-space correlation function for the 2dFGRS 
computed in this way is shown in Fig. 2. 
The correlation-function results display very clearly
the two signatures of redshift-space distortions discussed
above. The `fingers of God' from small-scale random
velocities are very clear, as indeed has been the case
from the first redshift surveys\nref{3}.
However, this is the first time that the 
detailed signature of large-scale
flattening from coherent infall has been seen with 
high signal-to-noise.

\noindent
{\big Quantifying redshift-space distortions}\eol
The large-scale flattening of the correlation function
may be quantified by measuring the quadrupole moment
of $\xi(\sigma, \pi)$ as a function of radius.
A negative quadrupole moment implies flattening, whereas
the finger-of-God distortion tends to yield a
positive quadrupole moment.
Fig. 3 shows that the quadrupole-to-monopole ratio is
positive on small scales, but that it falls with
separation, becoming progressively more negative up to
the largest separations at which it can
be reliably measured. This arises partly because the
underlying power spectrum is not a simple power law
function of scale, so that the peculiar
velocities have a different effect at different radii.
By integrating over the correlation function, it is possible to construct
quantities in which this effect is eliminated.
We shall not do this here, firstly because it seems
desirable to keep the initial analysis as direct
as possible. More importantly, finger-of-God smearing is
a significant correction that will also cause the flattening
to depend on radius. We therefore have to fit the data with
a two-parameter model, described in the caption to Fig. 2. 
The parameters are $\beta$ and a
measure of the size of the random dispersion in the relative
velocities of galaxies, $\sigma_p$. In practice, $\sigma_p$ plays the
role of an empirical fitting parameter to describe the scale on which
the distortions approach the linear-theory predictions.
It therefore also incorporates other possible effects, such as a 
scale dependence of bias.

\fig{3}
{The flattening of the redshift-space correlation function is
quantified by the quadrupole-to-monopole ratio,
$\xi_2/\xi_0$.
This quantity is positive where
fingers-of-God distortion dominates, and is negative
where coherent infall dominates.
The solid lines show model predictions for
$\beta=0.3$ 0.4 and 0.5, with 
a pairwise velocity dispersion of $\sigma_p=400\kms$
(solid lines), plus $\beta=0.4$  with $\sigma_p=300$
and $500\kms$ (dashed lines). The $\xi_2/\xi_0$ ratio
becomes more negative as $\beta$ increases and as $\sigma_p$ decreases.
At large radii, the effects of fingers-of-God become
relatively small, and values of $\beta\simeq 0.4$
are clearly appropriate.\eol
\strut\hglue\parindent
The multipole moments of the correlation function are defined as
$\xi_\ell(r) \equiv (2\ell+1)/2 \int_{-1}^1 \xi(\sigma=r\sin\theta,
\pi=r\cos\theta)\; P_\ell(\cos\theta)\; d\cos\theta.$
In linear theory, the quadrupole-to-monopole ratio is given\nref{29} by
$\xi_2 / \xi_0 = f(n) \,
(4\beta/3 + 4\beta^2/7) \,/\,
(1 + 2\beta/3 + \beta^2/5).$
Here $f(n)=(3+n)/n$, where $n$ is the power-spectrum index
of the density fluctuations: $\xi \propto r^{-(3+n)}$. 
In practice, nonlinear effects mean that this ratio is a function of scale.
We model this by
using the real-space correlation function estimated
from the APM survey\nref{13,16}, plus the model for nonlinear
finger-of-God smearing given in the caption to Fig. 2.
}
{70}
{\epsfbox[60 211 509 633]{xisigpiquad.eps}}

\fig{4}
{Likelihood contours for $\beta$ and the  
fingers-of-God smearing parameter $\sigma_p$,
based on the data in Fig. 3 
(considering $8 \mpcoh < r < 25 \mpcoh$).
These are plotted at the usual positions for
one-parameter confidence of 68\% (shaded region), and
two-parameter confidence of 68\%, 95\% and
99\% (i.e. $\Delta\chi^2 = 1, 2.3, 6.0, 9.2$).
The maximum-likelihood solution is
$\beta=0.43$ and $\sigma_p = 385 \kms$.
The value for the large-scale pairwise dispersion is in
reasonable agreement with previously suggested values\nref{30};
however, for the present analysis $\sigma_p$ is an uninteresting parameter.
If we marginalize over $\sigma_p$ (i.e. integrate
over $\sigma_p$, treating the likelihood as a
probability distribution), the final
estimate of $\beta$ and its rms uncertainty is
$\beta = 0.43 \pm 0.07$.\eol
\strut\hglue\parindent
We believe that this result is robust, in the sense that
systematic errors in the modelling are smaller than the
random errors. We have tried assuming that the power
spectrum for $k<0.1\hompc$ has the shape of a
$\Omega=0.3$ $\Lambda$CDM model, rather than the APM
measurement; this has a very small effect. A more serious
issue is whether the pairwise velocity dispersion of galaxies may
depend strongly on separation, as is found for mass
particles in numerical simulations\nref{31}. 
Assuming that the pairwise velocity dispersion $\sigma_p$
rises to twice its large-scale value below $1 \mpcoh$
reduces the best-fit $\beta$ by 0.04. This correction is small
because our analysis excludes the nonlinear data at $r<8 \mpcoh$.
}
{70}
{\epsfbox[60 211 509 633]{xisigpiconf.eps}}

The results for the quadrupole-to-monopole ratio are shown
in Fig. 3, which shows the average of the estimates for the
NGP and SGP slices. The difference between the NGP and
SGP allows an estimate of the errors to be made:
these slices are independent samples for the present analysis of
clustering on relatively small scales.
For model fitting, it is necessary to know
the correlation between the values at different $r$.
A simple way of addressing this is to determine
the effective number of degrees of freedom from the
value of $\chi^2$ for the best-fitting model.
A more sophisticated approach is to generate realizations
of $\xi(\sigma,\pi)$, and construct the required
covariance matrix directly. One way of achieving this
is to analyze large numbers of mock surveys drawn from 
numerical simulations\nref{32}. A more convenient method is to generate
direct realizations  of the redshift-space power spectrum,
using Gaussian fluctuations on large scales, but allowing
for enhanced variance in power on non-linear scales\nref{33,34,35}.
In practice, the likelihood contours resulting from this
approach agree well with those from the simple approach,
and we are confident that the resulting errors on $\beta$ are realistic.
These contours are shown in Fig. 4, and show that there is
a degree of correlation between the preferred values of
$\beta$ and $\sigma_p$, as expected. For present purposes,
$\sigma_p$ is an uninteresting parameter, so we marginalize over it
to obtain the following estimate of $\beta$ and its rms uncertainty:
$$
\beta = 0.43 \pm 0.07.
$$

This result is the first precise 
determination of $\beta$ from redshift-space
distortions. The best previous studies\nref{18,25,26} have
achieved no more than a detection of the effect at the $3\sigma$ level.
Before discussing the implications
of our result, we should therefore consider some possible
small systematic corrections that have been unimportant in earlier work.
First, the Kaiser analysis applies only in the small-angle
approximation, and in principle corrections might be needed
for wide-angle surveys such as ours\nref{36}.
However, with our weighting scheme, the mean angular
separation of pairs with spatial separations $< 30 \mpcoh$
is only $2.5^\circ$, so this is not a concern.
There is potentially a significant correction for luminosity effects.
The optimal weighting means that our mean
luminosity is high: it is approximately $M_{b_{\scriptscriptstyle\rm J}}=-20.3$, or
1.9 times the characteristic luminosity, $L^*$,  of the overall
galaxy population\nref{37}. 
A number of studies\nref{38} have suggested that the strength of
galaxy clustering increases with luminosity, with
an effective bias that can be fitted by $b/b^* = 0.7 + 0.3(L/L^*)$.
This effect has been controversial\nref{39}, but the 2dFGRS
dataset favours a similar luminosity dependence, as will be described elsewhere.
We therefore expect that $\beta$ for $L^*$ galaxies will exceed our directly measured
figure. Applying a correction using the given formula for $b(L)$,
we deduce
$$
\beta(L=L^*) =  0.54 \pm 0.09.
$$
Finally, the 2dFGRS has a median redshift of 0.11. With weighting,
the mean redshift in the present analysis is $\bar z =0.17$,
and our measurement should be interpreted as $\beta$ at
this epoch. The extrapolation to $z=0$ is model-dependent,
but probably does not introduce a significant change\nref{40}.

\fig{5}
{The dimensionless matter power spectrum at zero
redshift, $\Delta^2(k)$, as predicted from the
allowed range of models that fit the microwave-background
anisotropy data,
plus the assumption that $H_0 = 70 \kmsmpc \pm 10$\%. The solid line shows the
best-fit model\nref{41} (power-spectrum index $n=1.01$, and
density parameters in baryons, CDM, and vacuum of
respectively 0.065, 0.285, 0.760). The effects of
nonlinear evolution have been included\nref{42}.
The shaded band shows the $1\sigma$ variation around this
model allowed by the CMB data.
The solid points are the real-space power spectrum measured for APM galaxies.
The clear conclusion is that APM galaxies are consistent
with being essentially unbiased tracers of the mass on large scales.
Since the CMB data also constrain the range of
$\Omega$, this allows $\beta$ to be predicted.
}
{70}
{\epsfbox[60 211 509 633]{cmbpred_pk.eps}}

\noindent
{\big Consistency with microwave-background anisotropies}\eol
Our results are significant in a number of
ways. First, we have verified in some detail that the
pattern of redshift-space distortions associated with
the gravitationally-driven growth of clustering exists
as predicted. Although gravitational instability is 
well established as the standard model for the formation
of large-scale structure, it is an important landmark to
have verified such a characteristic feature of the theory.
Extracting the full cosmological implications of our measurement
of $\Omega^{0.6}/b$ requires us to know the bias parameter
in order to determine $\Omega$. 
For example, our measurement implies $\Omega=0.36 \pm 0.10$ if
$L^*$ galaxies are unbiased, but it is difficult to justify
such an assumption.
In principle, the details of the clustering pattern in
the nonlinear regime allow the $\Omega - b$ degeneracy to be broken,
yielding a direct determination of the degree of bias\nref{43}.
We expect to pursue this approach using the 2dFGRS.
For the present, however, it is interesting to use an
independent approach. Observations of anisotropies in the 
cosmic microwave background (CMB)
can in principle measure 
almost all the cosmological parameters, and current
small-scale anisotropy results are starting to tighten
the constraints. In a recent analysis\nref{41}, best-fitting values
for the densities in collisionless matter ($c$), baryons ($b$),
and vacuum ($v$) have been obtained:
$\Omega_c+\Omega_b+\Omega_v=1.11\pm0.07$,
$\Omega_c h^2=0.14 \pm 0.06$, $\Omega_b h^2 = 0.032\pm 0.005$,
together with a power-spectrum index $n=1.01\pm0.09$.
Our result for $\beta$ gives an independent test of this picture, as follows.

The only parameter left undetermined by the CMB data is the Hubble constant, $h$.
Recent work\nref{44,45} indicates that this is now determined to an
rms accuracy of 10\%,
and we adopt a central value of $h=0.70$. This completes
the cosmological model, requiring a total matter density parameter
$\Omega\equiv \Omega_c+\Omega_b=0.35 \pm 0.14$.
It is then possible to use the parameter limits from the 
CMB to predict a conservative range for 
the mass power spectrum at $z=0$, which is shown in Fig. 5.
A remarkable feature of this plot is that the mass power spectrum
appears to be in good agreement with the clustering observed in
the APM survey. For each model allowed by the CMB, we can predict
both $b$ (from the ratio of galaxy and mass spectra) and also
$\beta$ (since a given CMB model specifies $\Omega$).
In practice, we determine $b$ by determining the mean ratio
of power spectra over the range $0.02 < k < 0.1 \hompc$, where the
APM measurement is robust and where scale-dependent bias
and nonlinearities should be unimportant.
Considering the allowed range of models, we then obtain the prediction
$\beta_{\rm\scriptscriptstyle CMB+APM}  = 0.57 \pm 0.17$.
A flux-limited survey such as the APM will have a mean luminosity
close to $L^*$, so the appropriate comparison is with
the 2dFGRS corrected figure of $\beta = 0.54\pm0.09$ for $L^*$ galaxies. 
These numbers are in very close agreement.

This analysis of galaxy clustering in the 2dFGRS thus 
gives strong support to the simplest picture of cosmological structure
formation, in which the primary mechanism is gravitational
instability in a sea of collisionless dark matter.
We have shown that the fluctuations seen in the CMB (which
measure structure at a redshift $z\simeq 1100$) can be extrapolated
to the present to predict the peculiar velocities that distort
redshift-space clustering. The agreement between this
extrapolation and direct observations from the 2dFGRS is a
remarkable and highly non-trivial test of the basic model.
The precision of data in both areas should improve rapidly,
and the use of $\beta$ as a meeting ground between studies
of the CMB and large-scale structure  will  undoubtedly
lead to more demanding tests of the theory in years to come.
For the present, we can say that there is 
complete consistency between clustering in the 2dFGRS 
and the emerging `standard model' of cosmology:
a spatially flat, vacuum-dominated universe with density
parameter $\Omega \simeq 0.3$.

\bigskip
\ref{1} Hubble E.P., The Distribution of Extra-Galactic Nebulae, {\it Astrophys. J.}, 79, 8--76 (1934)
\ref{2} Kirshner R.P., Oemler A., Schechter P.L., Shectman S.A., A million cubic megaparsec void in Bootes, {\it Astrophys. J.}, 248, L57--L60 (1981)
\ref{3} Davis M., Peebles, P.J.E., A survey of galaxy redshifts. V - The two-point position and velocity correlations, {\it Astrophys. J.}, 267, 465--482 (1983)
\ref{4} Bean A.J., Ellis R.S., Shanks T., Efstathiou G., Peterson B.A., A complete galaxy redshift sample. I - The peculiar velocities between galaxy pairs and
                               the mean mass density of the Universe, {\it Mon. Not. R. astr. Soc.}, 205, 605--624 (1983)
\ref{5} de Lapparent V. Geller M.J., Huchra J.P., A slice of the universe, {\it Astrophys. J.}, 302, L1--L5 (1986)
\ref{6} Kaiser N., A sparse-sampling strategy for the estimation of large-scale clustering from redshift surveys, {\it Mon. Not. R. astr. Soc.}, 219, 785--790 (1986)% sparse-sampling & J_3 wt
\ref{7} Saunders W., Frenk C., Rowan-Robinson M., Efstathiou G.,  Lawrence A., Kaiser N., Ellis R., Crawford J., Xia X.-Y., Parry I., The density field of the local universe, {\it Nature}, 349, 32--38 (1991)
\ref{8} Shectman S.A., Landy S.D., Oemler A., Tucker D.L., Lin H., Kirshner R.P., Schechter P.L., The Las Campanas Redshift Survey, {\it Astrophys. J.}, 470, 172--188 (1996)
\ref{9} For details of the 2dF instrument, see {\tt http://www.aao.gov.au/2df/}
\ref{10} Taylor K., Cannon R.D., Watson F.G., Anglo-Australian Telescope's 2dF Facility, Proc. SPIE, 2871, 145--149 (1997)
\ref{11} Colless M., First results from the 2dF Galaxy Redshift Survey, {\it Phil. Trans. R. Soc. Lond.} A, 357, 105--116 (1999)
\ref{12} For details of the current status of the 2dFGRS, see \hfill {\tt http://www.mso.anu.edu.au/2dFGRS/}
\ref{13} Maddox S. Efstathiou G., Sutherland W.J., The APM Galaxy Survey - III. an analysis of systematic errors in the angular correlation function and cosmological implications, {\it Mon. Not. R. astr. Soc.}, 283, 1227--1263 (1996)
\ref{14} Ballinger W.E., Peacock J.A., Heavens A.F., Measuring the cosmological constant with redshift surveys, {\it Mon. Not. R. astr. Soc.}, 282, 877--888 (1996)
\ref{15} Hatton S., Cole S., Modelling the redshift-space distortion of galaxy clustering, {\it Mon. Not. R. astr. Soc.}, 296, 10--20 (1998)
\ref{16} Baugh C.M., Efstathiou G., The three-dimensional power spectrum measured from the APM Galaxy Survey-2. Use of the two-dimensional power spectrum, {\it Mon. Not. R. astr. Soc.}, 267, 323--332 (1994)
\ref{17} Kaiser N., Clustering in real space and in redshift space, {\it Mon. Not. R. astr. Soc.}, 227, 1--21 (1987)
\ref{18} Hamilton A.J.S., Tegmark M., Padmanabhan N., 2000, {\it Mon. Not. R. astr. Soc.}, 317, L23--L27
\ref{19} Pen U.-L., Reconstructing Nonlinear Stochastic Bias from Velocity Space Distortions, {\it Astrophys. J.}, 504, 601--606, (1998)
\ref{20} Dekel A., Lahav O., Stochastic Nonlinear Galaxy Biasing, {\it Astrophys. J.}, 520, 24--34, (1999)
\ref{21} Benson A.J., Cole S., Frenk C.S., Baugh C.M., Lacey C.G., The nature of galaxy bias and clustering, {\it Mon. Not. R. astr. Soc.}, 311, 793--808 (2000)
\ref{22} Kauffmann G., Nusser A., Steinmetz M., Galaxy formation and large-scale bias, {\it Mon. Not. R. astr. Soc.}, 286, 795--811 (1997)
\ref{23} Strauss M.A., Willick J.A., The density and peculiar velocity fields of nearby galaxies, {\it Phys. Reports}, 261, 271--431 (1995)
\ref{24} Hamilton A.J.S., Linear Redshift Distortions: a Review, in {\it The Evolving Universe\/}, Kluwer Astrophysics and space science library, 231, 185 (1998) (astro-ph/9708102)
\ref{25} Taylor A.N., Ballinger W.E., Heavens A.F., Tadros H., Application of Data Compression Methods to the Redshift-space distortions of the PSCz galaxy catalogue, {\it Mon. Not. R. astr. Soc.} in press (2000) (astro-ph/0007048)
\ref{26} Outram  P.J., Hoyle F., Shanks T., The Durham/UKST Galaxy Redshift Survey - VII. Redshift-space distortions in the power spectrum, {\it Mon. Not. R. astr. Soc.} in press (2000) (astro-ph/0009387)
\ref{27} Hamilton A.J.S., Toward Better Ways to Measure the Galaxy Correlation Function, {\it Astrophys. J.}, 417, 19--35 % xi estimators (1993)
\ref{28} Landy S.D., Szalay A.S., Bias and variance of angular correlation functions, {\it Astrophys. J.}, 412, 64--71 % xi estimators (1993)
\ref{29} Hamilton A.J.S., Measuring Omega and the real correlation function from the redshift correlation function, {\it Astrophys. J.}, 385, L5--L8 (1992) % xi_2/xi_0
\ref{30} Jing Y.P., Mo H.J., B\"orner G., Spatial Correlation Function and Pairwise Velocity Dispersion of Galaxies: Cold Dark Matter Models versus the Las Campanas Survey, {\it Astrophys. J.}, 494, 1--12 (1998)
\ref{31} Jenkins A.R. et al., Evolution of Structure in Cold Dark Matter Universes, {\it Astrophys. J.}, 499, 20--40 (1998)
\ref{32} Cole S., Hatton S., Weinberg D.H., Frenk C.S., Mock 2dF and SDSS galaxy redshift surveys, {\it Mon. Not. R. astr. Soc.}, 300, 945--966 (1998)
\ref{33} Feldman H.A., Kaiser N., Peacock J.A., Power-spectrum analysis of three-dimensional redshift surveys, {\it Astrophys. J.}, 426, 23--37 (1994)
\ref{34} Meiksin A.A., White M., The growth of correlations in the matter power spectrum, {\it Mon. Not. R. astr. Soc.}, 308, 1179--1184 (1999)
\ref{35} Scoccimarro R., Zaldarriaga M., Hui L., Power Spectrum Correlations Induced by Nonlinear Clustering, {\it Astrophys. J.}, 527, 1--15 (1999)
\ref{36} Szalay A.S., Matsubara T., Landy S.D., Redshift-Space Distortions of the Correlation Function in Wide-Angle Galaxy Surveys, {\it Astrophys. J.}, 498, L1--L4 (1998)
\ref{37} Folkes S.J., et al., The 2dF Galaxy Redshift Survey: spectral types and luminosity functions, {\it Mon. Not. R. astr. Soc.}, 308, 459--472 (1999)
\ref{38} Benoist C., Maurogordato S., da Costa L.N., Cappi A., Schaeffer R., Biasing in the Galaxy Distribution, {\it Astrophys. J.}, 472, 452--459 (1996)
\ref{39} Loveday J., Maddox S.J., Efstathiou G., Peterson B.A., The Stromlo-APM redshift survey. 2: Variation of galaxy clustering with morphology and luminosity, {\it Astrophys. J.}, 442, 457--468 (1995)
\ref{40} Carlberg R.G., Yee H.K.C., Morris S.L., Lin H., Hall P.B., Patton D., Sawicki M., Shepherd C.W., Galaxy Clustering Evolution in the CNOC2 High-Luminosity Sample, {\it Astrophys. J.}, 542, 57--67 (2000)
\ref{41} Jaffe A., et al., Cosmology from Maxima-1, Boomerang and COBE/DMR CMB Observations, astro-ph/0007333 (2000)
\ref{42} Peacock J.A., Dodds S.J., Non-linear evolution of cosmological power spectra, {\it Mon. Not. R. astr. Soc.}, 280, L19--L26 (1996)
\ref{43} Verde L., Heavens A.F., Matarrese S., Moscardini L., Large-scale bias in the Universe - II. Redshift-space bispectrum, {\it Mon. Not. R. astr. Soc.}, 300, 747--756 (1998)
\ref{44} Mould J.R. et al., The Hubble Space Telescope Key Project on the Extragalactic Distance Scale. XXVIII. Combining the Constraints on the Hubble Constant, {\it Astrophys. J.}, 529, 786--794 (2000)
\ref{45} Freedman W.L. et al., Final Results from the Hubble Space Telescope Key Project to Measure the Hubble Constant, astro-ph/0012376 (2000)

\medskip
\noindent
{\sc acknowledgements}. The 2dF Galaxy Redshift Survey was made possible through
the dedicated efforts of the staff of the Anglo-Australian Observatory,
both in creating the 2dF instrument and in supporting it on the telescope.

\bye